\documentclass[11pt,a4paper]{article}
\usepackage{cite}
\usepackage{appendix}
\usepackage[T1]{fontenc}
\usepackage{LibreBodoni}
\usepackage{amsmath, amsthm, mathtools, url, amssymb, upgreek, slashed, times, color, enumerate, mathrsfs}
\usepackage{enumitem}
\usepackage{soul}
\usepackage{tabto}
\usepackage{lipsum}
\newcommand{\inn}{\!\in \!}
\parskip=\baselineskip
\newcommand\blfootnote[1]{%
  \begingroup
  \renewcommand\thefootnote{}\footnote{#1}%
  \addtocounter{footnote}{-1}%
  \endgroup
}

\begin{document}

\begin{titlepage}
\vskip 1.5in
\begin{center}
{
\bf
\Large{Infinite Permutation Groups and the \\ Origin of Quantum Mechanics}}
\vskip 0.5cm 
{Pavlos Kazakopoulos and Georgios Regkas} 
\vskip 0.05in 
{
\small{
\textit{Physics Dept., University of Athens}
\vskip -.4cm
{\textit{Athens, 157 84, Greece}}
}
}
\end{center}
\vskip 0.5in
\baselineskip 16pt

\begin{abstract}
We propose an interpretation for the meets and joins in the lattice of experimental propositions of a physical theory, answering a question of Birkhoff and von Neumann in \cite{logic_of_qm}. When the lattice is atomistic, it is isomorphic to the lattice of definably closed sets of a finitary relational structure in First Order Logic. In terms of mapping experimental propositions to subsets of the atomic phase space, the meet corresponds to set intersection, while the join is the definable closure of set union. The relational structure is defined by the action of the lattice automorphism group on the atomic layer. Examining this correspondence between physical theories and infinite group actions, we show that the automorphism group must belong to a family of permutation groups known as geometric Jordan groups. We then use the classification theorem for Jordan groups to argue that the combined requirements of probability and atomicism leave uncountably infinite Steiner $2$-systems (of which projective spaces are standard examples) as the sole class of options for generating the lattice of particle Quantum Mechanics.

\end{abstract}

\blfootnote{*Correspondence: pkazakop@phys.uoa.gr}

\end{titlepage}

\section{Introduction}

At the beginning of the 20th century, human understanding of nature was revolutionised by the discovery of Quantum Mechanics. The first axiomatization of particle quantum mechanics was given by von Neumann in 1932 in the language of complex Hilbert spaces \cite{von2018mathematical}, and it remains the standard today. In the decade that followed, two interrelated research directions were opened up for achieving a deeper comprehension of the mathematical and conceptual foundations of the theory. In \cite{logic_of_qm}, Birkhoff and von Neumann used lattice theory to describe and compare classical and quantum mechanics in a common language. In \cite{murray1936rings,murray1937rings,neumann1940rings,murray1943rings}, Murray and von Neumann created the theory of rings of operators, or von Neumann algebras, a very powerful mathematical framework that can express essentially all known theories of Physics. The projection lattices of some type $II$ and $III$ von Neumann algebras are non-atomic physical lattices that arise in Quantum Field Theory \cite{araki1964type,haag2012local,witten2022gravity,chandrasekaran2023algebra}. 

The lattice description has come to be known as Quantum Logic. The idea is that in theories of Physics physical phenomena can be labeled in a manner that encodes an implication relation between them. In a given theoretical description, we say that statement $p$ about the system implies statement $q$ about the system, if, whenever the theory interprets an observation as assigning True to $p$, it also necessarily interprets it as assigning True to $q$. The operational meaning is that statement $p$ describes a more specific, or precise, observation\footnote{For example, assume that in some experiment on the system we measure the value of an observable $O$ (e.g. the energy) to be in a certain interval of real numbers $[a, b]$ (in some unit system). Then this measurement renders True the descriptive statement: ``The value of $O$ is in $[a, b]$'' as well as any statement ``The value of $O$ is in $I$'' where $I$ is any set of reals containing $[a, b]$. This is of  course a trivial example but it becomes less so when more than one observables are involved, due to non-commutativity of observables in quantum mechanics.} of the system than $q$. This type of implication can be abstracted to a partial order relation between $p$ and $q$, written $p \leq q$. The set of statements is turned into a partially ordered set $\mathcal{P}$, via the truth assignments associated with experimental outcomes. Some reasonable additional requirements on $\mathcal{P}$ promote it to a complete lattice $\mathcal{L}$. If a certain experimental outcome is interpreted as assigning True to a subset of elements of $\mathcal{L}$, then the infimum (meet) of this subset can be used to label that particular outcome. In this manner the elements of $\mathcal{L}$ label classes of experimental outcomes that are implicationally equivalent, and make the implication structure manifest.

The existence of an implication structure in the statements of the theory, resulting from the theoretical interpretation of experimental outcomes, is reminiscent of implication between propositions in propositional logic (zeroth-order logic). The analogy is supported by the observation that $\mathcal{L}$ is actually a distributive lattice in the case of classical mechanics. It can then always be lattice-embedded into a Boolean algebra, which by Stone's representation theorem \cite{stone1936theory} is isomorphic to an algebra of sets. The meet and join operations on the lattice of classical mechanics work exactly like the AND/OR logical connectives in propositional logic, represented by set intersection and union in an algebra of sets. Birkhoff and von Neumann coined the term ``experimental propositions'' for the elements of $\mathcal{L}$. Resemblances between physics and propositional logic end there however. As Birkhoff and von Neumann show, non-distributivity of $\mathcal{L}$ in quantum mechanics precludes an interpretation of meets and joins as classical logical operators. In the suggested questions of \cite{logic_of_qm}, they ask for the interpretation of meets and joins in an arbitrary physical theory. Here we propose an answer to this question. We show that a proper account and interpretation of $\mathcal{L}$ can be given in Group Theory and First Order Logic (FOL). The elements of an atomic physical lattice $\mathcal{L}$ are definably closed sets of a relational structure $\mathcal{R}$ in FOL, whose domain is the set of lattice atoms $A$. The meet $p \wedge q$ and join operations $p \vee q$ correspond to set intersection $p \cap q$ and the definable closure of set union $\textbf{dcl}(p \cup q)$, respectively. This provides a mapping from the labels of phenomena to the complete lattice of definably closed sets of atoms, in which implication is represented by set inclusion. The first order relational structure $\mathcal{R}$ is itself determined by the automorphisms of $\mathcal{L}$, through their action on the atomic layer. Meets and joins in $\mathcal{L}$ can be calculated directly from the automorphism group action $(G, A)$, where $A$ are the atoms of $\mathcal{L}$. Operationally, the meets and joins of $\mathcal{L}$ are rules for producing labels for experimental outcomes that make evident their implication structure, i.e. whether a measurement is more specific than another. The labels are sets of atomic names, and implication between experimental outcomes is manifest as set inclusion between the corresponding labels. The atomic names of the theory are assigned to ``maximal observations'' \cite{logic_of_qm} of the system.

The mathematical assumption that makes this representation possible is the reconstructibility of $\mathcal{L}$, up to a closure operation, from the action of its own group of automorphisms $(G, \mathcal{L})$. It is a strong assumption, by no means true for generic lattices, but in the context of physics it is a natural one. The elements of $\mathcal{L}$ are labels of experimental outcomes. A proper subset $S \! \subset \! \mathcal{L}$ in which all elements are implied only by other elements of $S$ and also has a maximum element, can be interpreted as a subsystem of the original system. Operationally, it corresponds to a sector of measurements refining in precision a given experimental outcome (the maximum element of $S$). The lower cone $l_{\downarrow} \! \subseteq \! \mathcal{L}$ of an element $l$ consists of all elements that imply it: $l_{\downarrow}=\{l' \; | \; l' \leq l\}$. The subsystems of the original system are precisely the lower cones of $\mathcal{L}$ under this definition. Labeling the phenomena in any subsystem should not ineluctably fix the label of any other outcome outside that particular subsystem. Allowed relabelings are realised by lattice automorphisms. Therefore, the pointwise stabiliser $G_{l_{\downarrow}}$ of the lower cone of any lattice element $l \inn \mathcal{L}$ in $(G, \mathcal{L})$ must be able to move all other lattice elements, i.e. $l'  \notin l_{\downarrow} \!\! \implies \!\! G_{l_{\downarrow}} \nleqslant G_{l'}$ which implies that $l \neq l' \!\! \implies \!\! G_{l_{\downarrow}} \neq G_{l'_{\downarrow}}$. In an atomistic lattice, lower cones correspond uniquely to sets of atoms. This guarantees the existence of an order-preserving injection $\boldsymbol{\pi}: \mathcal{L} \rightarrow \mathcal{L}_{\mathrm{fix}}$ into the lattice of fixets of the action $(G, A)$ induced on the set of atoms $A$ by $(G, \mathcal{L})$. The fixsets are the subsets of $A$ that can be pointwise stabilised by its automorphisms. Every subgroup $H$ of the Symmetric Group, $Sym(A)$, defines a first order relational structure $\mathcal{R}$ (\!\cite{hodges1993model}, Thm. 4.1.4) on the domain $A$ whose automorphism group, $G = Aut(\mathcal{R})$, is the topological closure\footnote{In the pointwise convergence topology, where the closure of a subgroup of permutations is all permutations that agree with these on all finite sets. The closure is also a group.} of $H$. This is done by taking the orbits of its action on $A^n,\; n \inn \mathbb{N}$, as the defining relations of $\mathcal{R}$. The fixsets of $(G, A)$ are the definably closed sets\footnote{This is not always the case (see \cite{hodges1993model}, ch. 4) but in known physically relevant structures it is.} of $\mathcal{R}$ which form a complete lattice, $\mathcal{L}_{\mathrm{dcl}}$. Since $\mathcal{L}_{\mathrm{fix}} \cong \mathcal{L}_{\mathrm{dcl}}$, there is an order preserving injection $\boldsymbol{\rho}: \mathcal{L} \rightarrow \mathcal{L}_{\mathrm{dcl}}$. The image of $\boldsymbol{\rho}$ is always a closure operation $\mathbf{cl}: \mathcal{L}_{\mathrm{dcl}} \rightarrow \mathcal{L}_{\mathrm{dcl}}$. In almost all cases of interest in Physics, $\mathbf{cl}$ is the identity, making $\boldsymbol{\rho}$ a lattice isomorphism $\mathcal{L} \cong \mathcal{L}_{\mathrm{dcl}}$. But there is one example where it is not so, the projection lattice of the von Neumann $I_2$ factor. In this case the theory exhibits special features, such as admitting a classical hidden variable model as has been shown by explicit construction in \cite{kochen1967problem}. Non-atomic physical lattices appear as sublattices of $\mathcal{L}_{\mathrm{dcl}}$. They include the lattice of non null measurable sets of the phase space in classical mechanics and the projection lattices of type $II$-$III$ von Neumann factors in quantum field theory.

Borrowing a term from computer science, we can call $\mathcal{R}$ the namespace of the system. It is a relational structure on a domain $A$ of elementary labels, or atomic names, from which all the labels of experimental outcomes are produced as definably closed subsets in $\mathcal{R}$. The action $(G, A)$ governs the operation of definable closure that produces the lattice elements. Accordingly, the permutation group $G \leqslant Sym(A)$ representing the action can be called the theory's naming group. Taking the definable closure of a subset of the domain means including all those points that can be defined (that is, uniquely specified) by the selected points using the structure's relations. For example, in a vector space the definable closure of any set of vectors $\{\psi_{I}\}$ is the linear subspace they define, $\mathbf{span}(\{\psi_{I}\})$, as all vectors contained in the subspace can be reached from the original selection through the collinearity relations\footnote{One way to define a vector space $V$ over a field $F$ as a relational structure in FOL is to consider the language $\{0, +, -, \mathcal{V}_{\alpha}\}$ where the first three entries account for the abelian group structure of the vector space and $\mathcal{V}_{\alpha}, \; \alpha \inn F$ are binary relations expressing multiplication by $\alpha$, i.e. $\mathcal{V}_{\alpha}(\psi', \psi)$ iff $\, \psi'= \alpha \psi \,$ in common notation. We can also write, in function notation, $\psi'=\mathcal{V}_{\alpha}(\psi)$ Then, given any set of vectors $\{\psi_{I}\}$ their definable closure will be all possible finite linear combinations of elements in the set, i.e. $\mathbf{dcl}(\{\psi_{I}\})=\mathbf{span}(\{\psi_{I}\})$. If we take the equivalence classes mod $\mathcal{V}_{\alpha}$ in $V \! - \! \{0\}$ we obtain the projective space $P(V)$ and the definably closed sets are the corresponding projective subspaces. This is the superposition principle in Quantum Mechanics.} defining the vector space. In physically relevant structures we shall see that they are precisely the fixsets of the group action $(G, A)$. We shall use the terms fixset and definably closed set interchangeably.

As a theoretical object, in atomic theories $\mathcal{R}$ is a precursor to the phase space $\Phi$ of the physical system. By this we mean that it has the same domain A, but more symmetries than the phase space. $\Phi$ is derived from $\mathcal{R}$ by adding relations that express statistical or dynamical properties of the system that are not captured by the implication structure encoded in $\mathcal{L}$. The naming group is a proper supergroup of the automorphism group of the phase space, for example the Hamiltonian symplectomorphism groups of symplectic manifolds $Ham(\Phi)$ in Hamiltonian classical mechanics, or the projective unitary group $PU(n, \mathbb{C}), \;2 \leq n \leq \aleph_0 $ of projective Hilbert space in quantum mechanics. In classical physics the naming group is the symmetric group $Sym(A)$, acting on a namespace $A$ of cardinality $2^{\aleph_0}$ for particle mechanics, or $2^{2^{\aleph_0}}$ for classical field theory (since the latter concerns possible values of the fields, which are a continuum, at every point of the space continuum). $Ham(\Phi)$ is trivially a subgroup of $Sym(A)$. The set of atoms $A$ of the namespace is the set of maximal observations, or pure states, of the system (all possible exact positions and momenta of all particles, or values of fields and field momenta at every point of space). For quantum mechanics the naming group is the Projective Linear group $PGL(n, \mathbb{C}), \;2 \leq n \leq \aleph_0 $, acting on the rays of a complex vector space of finite or countably infinite dimension. The rays of the vector space are the pure states, the atomic names of the namespace. In this case also, $PU(n, \mathbb{C}) < PGL(n, \mathbb{C})$.

The atomistic lattices for these theories are simply the fixset lattices of the corresponding permutation groups. For classical particle mechanics and field theory this is the powerset lattice of the domain of the namespace, $\mathscr{P}(A)$. Birkhoff and von Neumann refer to this maximalistic option as the ``naive interpretation'' of classical mechanics. They take the position that the infinitely precise measurements defining the atomic propositions (``maximal observations'') are experimentally unattainable, and so only statements placing the state of the system in some non-null Lebesgue measurable subset of the namespace are operationally meaningful. The non-atomic lattice of Lebesgue measurable mod Lebesgue-null subsets of the namespace is a sublattice of the fixset lattice of $Sym(A)$. In particle quantum mechanics the domain of the namespace is the set of pure states and the lattice of propositions is the fixset lattice of the defining representation of $PGL(n, \mathbb{C})$. This is the lattice of subspaces of the projective space, isomorphic to the lattice of closed linear subspaces of the corresponding Hilbert space. The case $n=2$ is special, as its fixset lattice has enhanced symmetry. It corresponds to the $I_2$ von Neumann factor and displays interesting properties that we shall discuss. Quantum Field Theory uses type $II$-$III$ von Neumann algebras for the local observables. Their projection lattices are complete non-atomic sublattices of the fixset lattice of $PGL(\aleph_0, \mathbb{C})$.

Namespaces and naming groups as listed above apply to all observers when the universe is causally connected. Taking causality explicitly into account in causally disconnected universes is more subtle. So far we have talked about a namespace associated to a given theoretical description of a physical system. Certain subsets of the namespace (viz. the definably closed sets of a first order relational structure, or equivalently the fixsets of its automorphism permutation group) are used as labels in the context of the theory for the phenomena exhibited by the system. When causality is accounted for, one recognises that this naming group can be consistently used only by the observers that are in bidirectional causal contact with the system. We refer to observers along maximal timelike worldlines, as in \cite{witten2022does}. Not all causally connected observers will apply the same label to a given phenomenon. The reality of a physical event is ascertained in a physical theory by the existence of a group of symmetries - a subgroup of the naming group - that can relate any two distinct observations of that event (e.g. the Poincar\'e group for inertial observers in Minkowski space). Although two observers recording the same event will generally obtain different data (for example by using different reference frames), the group always allows a transformation between the two datasets. Group axioms ensure that these transformations are consistent for any set of observers with access to the event. In an arbitrary universe, not all observers are necessarily causally connected. Here, in order to focus on the main idea, we shall not incorporate causality explicitly in the description. 

The correspondence between lattices of physical theories and infinite permutation groups raises a new question: Which permutation groups can be used as naming groups for theories of Physics? Naming groups must be topologically closed, and those we encounter in established theories are of only two kinds: The Infinite Symmetric Group $Sym(X), |X|\inn \{2^{\aleph_0}, 2^{2^{\aleph_0}}\}$ in classical mechanics, and the Projective Linear group $PGL(n, \mathbb{C})$ of a (finite or countably infinite-dimensional) complex projective space for quantum mechanics. But closed subgroups of Infinite Symmetric Groups, even maximal ones, abound \cite{bergman2006closed,mitchell2011generating}. Which organizing principle could be applied to limit the possibilities? We have already applied a group-theoretic principle at the level of the lattice, as the requirement that in a physical lattice the labeling of a single subsystem cannot uniquely fix the label of any other phenomenon. This guarantees that the lattice can be reconstructed (modulo a closure) by its automorphisms, and the rest of the argument follows. From the perspective of the automorphism group action, we recognise it as a principle related to the transitivity properties of the action. We can investigate what happens if the condition is strengthened. Assume then that the labeling of the phenomena in any subsystem does not at all restrict the assignment of other atomic names to a maximal observation outside that subsystem. In terms of the action $(G, A)$: the pointwise stabiliser of any fixset is transitive on the atomic names not contained in the fixset. This principle might sound abstract, but its physical interpretation is uncomplicated. Apart from cosmological questions, any system in Physics can be considered a subsystem of a larger one. If, after fixing all the names in a given system $S$, one seeks to extend it to a given larger system $S_{super}$, the extension ought to be unique. That means that the pointwise stabiliser $G_{S}$ of the smaller system $S$ in the automorphism group of its new supersystem $S_{super}$ must be transitive on the maximal observations in the complement $S_{super} \setminus S$. For assume that it is not, i.e. in the action $(G_{S}, S_{super})$ maximal observations outside $S$ fall in more than one orbits. Then one would get inequivalent extensions of $S$ to $S_{super}$ depending on which of the orbits the label of the first maximal observation outside $S$ is chosen from, if one wished to preserve the labels already assigned to the smaller system. Applying the same reasoning in the downward direction, the pointwise stabiliser of any subsystem $S_{sub}$ of $S$ must have this property, i.e. be transitive on the maximal observations in $S \setminus S_{sub}$.

Besides $Sym(A)$, there is only one class of topologically closed permutation groups with this transitivity property: geometric Jordan groups. They are a subclass of Jordan groups, which have at least one fixset (of cardinality at least equal to the degree of transitivity) whose pointwise stabiliser is transitive on its complement. Geometric Jordan groups admit actions where the stabilisers of \emph{all} fixsets are transitive on their complements. Jordan groups have been studied since the 19th century \cite{jordan1871theoremes,neumann1985some,neumann2006concept} and were fully classified in the finite case in the 1980's \cite{kantor1985homogeneous,neumann1985some,cherlin1985N0,mcdonough1981infinite}. Infinite primitive Jordan groups of finite transitivity were also classified in the 1990's, in terms of the structures that they must preserve \cite{cameron1976transitivity,adeleke1996primitive,adeleke1996classification}. They are automorphism groups of structures that fall into two categories: One contains linear and semilinear orders and other weaker order-derived relations and appears in the infinite case only. The other is Steiner systems, which are geometric structures of points organised in blocks of equal cardinality. In a Steiner $k$-system ($k \geq 2$) every set of $k$ points uniquely defines a block (Euclidean space is a Steiner $2$-system where lines are the blocks). The Jordan groups preserving structures of the second category are called geometric \cite{bhattacharjee2006notes}, and examples appear in both the finite and infinite cases. We see that strengthening the initial transitivity assumption has drastic consequences, restricting admissible groups to automorphism groups of just one general type of structure, Steiner systems. The Projective Linear groups in any dimension over any field are geometric Jordan groups. They are automorphism groups acting on the rays of their corresponding vector spaces. This can be interpreted as action on a Steiner $2$-system where rays are the points and the 2-dimensional planes defined by pairs of rays are the blocks. The Symmetric Group, although has the property of being transitive on the complement of every fixset, does not fall into this classification, as it is highly transitive. It is, however, the only topologically closed highly transitive permutation group in all infinite cardinalities, and thus the only possible highly transitive automorphism group of a first order relational structure (a plain set of the given cardinality).

According to the classification theorem, a Jordan group preserving a Steiner $k$-system is $k$-transitive but not $(k+1)$-transitive. Steiner systems on infinite domains have not been classified. All infinite Steiner $k$-systems with Jordan automorphism group for $k \geq 3$ are homogeneous extensions of Steiner $2$-systems \cite{beutelspacher1994transfinite,adeleke1996classification}. Their corresponding fixset lattices do not contain additional information relative to the lattices of Steiner $2$-systems to which they reduce, except the number of extensions $k-2$. Because of this they exhibit undesirable physical properties. Infinite Steiner $2$-systems have not been classified, but the only examples known in uncountable cardinalities are very familiar: affine and projective spaces. The automorphism groups of these structures are similarly well-known: The Affine ($AFL$) and Projective ($PGL$) groups are geometric Jordan groups. Affine spaces cannot support atomic transition probabilities because the translation part of the group makes normalization impossible. Thus it seems that the only known type of first-order structure that satisfies the enhanced transitivity assumption and can (after its completion to a projective Hilbert space) be used to model atomic transition probabilities is that of a projective vector space, as is the case in Quantum Mechanics. This argument, to the degree that it constrains the possibilities for the naming group of a theory that combines atomicism with probability, is contingent upon the understanding of the possibilities for infinite Steiner $2$-systems with Jordan automorphism groups (see remark 5.8.4 in \cite{adeleke1996infinite}). The subject is, to the best of our knowledge, currently unexplored and so the argument is inconclusive in that sense. A better understanding of the typology of continuously infinite Steiner $2$-systems is needed to show exactly how restrictive is the transitivity property we assumed for the naming group. If it were found that more types of Steiner $2$-systems that satisfy this condition exist, then to further constrain the class of appropriate structures for probabilistic atomic theories one would seem to need a combinatorial extension of Gleason's theorem \cite{gleason1957measures}, or a weaker version of it, showing for which of these systems atomic probabilities are, if not uniquely defined, then at least possible to define.

\section{Preliminaries}

We recall some mathematical facts for the reader's convenience and to establish notation. A lattice is a partially ordered set (poset) in which every pair of elements has a supremum (join) and infimum (meet). So if $\mathcal{P}$ is a poset and $p, q \inn \mathcal{P}$ then their join and meet are defined respectively as: 
\begin{eqnarray}\nonumber
& p \vee q = sup\{p, q\}\\ \nonumber
& p \wedge q = inf\{p, q\}. \nonumber
\end{eqnarray}
These do not always exist as elements of $\mathcal{P}$, but when they do they are unique. $\mathcal{P}$ is called a lattice $\mathcal{L}$ when the meet and the join exist for every pair of elements (and thus for every finite set thereof). If they exist for any set of elements, $\mathcal{L}$ is called complete. A poset can always be order-embedded into a unique minimal complete lattice by the Dedekind-MacNeille completion. A lattice that has a minimum and a maximum element, denoted $\mathbf{0}$ and $\mathbf{1}$, is called bounded. A complete lattice is always bounded. The powerset $\mathscr{P}(X)$ of any set $X$ is a complete lattice with join and meet being set union and intersection. A closure operation on a lattice $\mathcal{L}$ is a map $\mathbf{cl}: \mathcal{L}\rightarrow \mathcal{L}$ that has the following properties for all $p, q \inn \mathcal{L}$:
\begin{enumerate}[label=\roman*), ref=\roman*, topsep=0pt]
\item $p \leq \mathbf{cl}(p)$ (extensive)
\item $p\leq q \implies \mathbf{cl}(p) \leq \mathbf{cl}(q)$ (increasing)
\item $\mathbf{cl}(\mathbf{cl}(p))=\mathbf{cl}(p)$ (idempotent)
\end{enumerate}
The image of a closure operation on a complete lattice is also a complete lattice. 
A lattice filter $f$ is a non-empty set of lattice elements such that:
\begin{enumerate}[label=\roman*), ref=\roman*,  topsep=0pt]
\item $\mathbf{0} \notin \! f$ (proper)
\item $p \inn f$ and  $p \leq q \implies q \inn f$ (upwards closed)
\item $p, q \inn f \implies p \wedge q \inn f$ (closed under meets)
\end{enumerate}
A lattice ultrafilter $u$ is a lattice filter that is maximal (not contained in any other proper filter).

A group action of a group $G$ on a set $X$, denoted $(G, X)$, is a function $\alpha: G\times X \rightarrow X$ that satisfies $\alpha(e, x)=\alpha(x)$ and $\alpha(g, \alpha(h, x))=\alpha(gh, x)$ where $x\inn X$, $g, h\inn G$ and $e$ is the identity element of $G$. We shall write $gx$ for $\alpha(g, x)$. For any given $g$, the map $x\rightarrow gx$ is a bijection of the set $X$ to itself. We call an action faithful if every element of $G$ induces a different such bijection. In that case, the action defines an isomorphism between $G$ and a subgroup of $Sym(X)$. Since the action of a group on itself by left multiplication is always faithful, such an isomorphism exists for every group (Cayley's theorem). The orbit $O_x$ of a point $x\inn X$ is the set of all points that can be reached by the action of $G$, i.e. $O_x=\{gx \; | \; g \inn G\}$. An action is transitive if it consists of a single orbit, or in other words if $\; \forall x,y\inn  X \; \exists g\inn G \; y=gx$. It is called $k$-transitive if the same is true for any two $k$-tuples of distinct points from $X$, and highly transitive if this is true for all finite $k$. An action is called primitive if it does not preserve any proper non-trivial equivalence relation on $X$ (an equivalence relation is proper if it has more than one class and non-trivial if there is a class with more than one element). A primitive action is transitive, but not vice-versa. A counterexample is the action of $GL(V)$ on $V \! - \! \{0\}$, a vector space minus the origin. It is clearly point-transitive, but preserves the rays setwise.

For $\Delta \subseteq X$, we denote the pointwise stabiliser of $\Delta$ by $G_{\Delta}$ and its setwise stabiliser by $G_{\{\Delta\}}$. For a singleton set $\{x\}$ the setwise and pointwise stabilisers coincide and we can simply write $G_x$. Note that there are generally more than one subgroups that stabilise a set $\Delta$ either setwise or pointwise. $G_{\{\Delta\}}$ and $G_{\Delta}$ denote the maximum subgroups of each family, which can always be uniquely defined as $G_{max} = \langle G_I \rangle$, where the angled brackets denote the generated group and $I$ indexes the subgroups inside a family. The subgroups of any group form a complete lattice $\mathcal{L}_G$, under the natural subgroup order relation, with $\vee G_I = \langle G_I \rangle$ and $\wedge G_I = \cap \, G_I$ for any collection of subgroups $\{G_I\}$. The trivial subgroup $\{e\}$ and the whole group $G$ are the minimum and maximum elements of $\mathcal{L}_G$. The pointwise stabilisers $G_{\Delta}$ are a complete sublattice $\mathcal{L}_{\mathrm{stab}}$ of the subgroup lattice $\mathcal{L}_G$. The setwise stabilisers on the other hand are only a subposet of $\mathcal{L}_G$. Not all $\Delta \subseteq X$ can be stabilised independently. In general, stabilizing $\Delta$ will fix additional points of $X$. The action $(G, X)$ defines therefore through pointwise stabilization a closure operation on $\mathscr{P}(X)$ mapping each $\Delta$ to the corresponding fixset $\Delta_{\mathrm{fix}} \equiv \{ x \inn X \; | \; gx\!=\!x \; \forall g \inn G_{\Delta} \}$. Importantly for our purposes, $\Delta_{\mathrm{fix}}$ form a complete lattice $\mathcal{L}_{\mathrm{fix}}$. Note that while $\mathcal{L}_{\mathrm{fix}}$ is a subposet of $\mathscr{P}(X)$, it is not a sublattice. The meets are still represented by set intersection, but the joins turn out differently. The definitions of both are derived from the structure of $\mathcal{L}_{\mathrm{stab}}$. Let:
\begin{equation}
    \mathbf{fix}: \mathcal{L}_{\mathrm{stab}} \rightarrow \mathcal{L}_{\mathrm{fix}}, \;\; \mathbf{fix}(G_{\Delta})=\Delta_{\mathrm{fix}}
\end{equation}
be the bijection producing from a pointwise stabiliser the corresponding fixset. Since $\Delta' \subseteq \Delta \Leftrightarrow G_{\Delta} \leq G_{\Delta'}$ the bijection is order-reversing. It is in fact a Galois duality \cite{hillman}. If $\{\Delta_{I}\}$ is any collection\footnote{The index $I$ is arbitrary. We shall omit the curly brackets around collections in equations when there is no ambiguity.} of fixsets and $\{G_{\Delta_{I}}\}$ their corresponding pointwise stabilisers, then:
\begin{subequations}
\begin{align}
& \wedge \! \Delta_{I} =  \mathbf{fix}(G_{\cap \Delta_{I}}) = \mathbf{fix}(\langle G_{\Delta_{I}}\rangle) = \cap \Delta_{I} \\ 
& \vee \! \Delta_{I} =  \mathbf{fix}(G_{\cup \Delta_{I}}) = \mathbf{fix}(\cap \, G_{\Delta_{I}}) \supseteq \cup \Delta_{I} 
\end{align}
\end{subequations}

Any intransitive action $(G, X)$ can be written as a subdirect product $\mathbf{Sub}(\prod_{I}(G_{I}, X_{I}))$ of transitive actions $(G_{I}, X_{I})$, where $\{X_{I}\}$ is a partition of $X$ (\cite{hall2018theory} sec. 5.5). Recall that a subdirect product of groups is a subgroup of their direct product that projects surjectively on each factor. So transitive actions are the building blocks of all actions. Every group implicitly contains all its possible transitive actions in its abstract algebraic definition via the multiplication table. For if $(G, X)$ is a transitive action, consider a point $x \inn X$ and let $G_x \leq G$ be the pointwise stabiliser of the singleton $\{x\}$. Since the action is assumed transitive, any $y\inn X$ can be reached from $x$. If $y=gx$ then the elements of $G$ that satisfy this relation are exactly those in the subset $gG_x$ of $G$. It is easily seen that this is a bijection between the points of $X$ and the points of the coset space $G/G_x$, with $G$ acting by left multiplication on the latter. The action is primitive if and only if $G_x$ is a maximal subgroup of $G$. The two group actions, on the original space and on the coset space, are isomorphic. Reversing these steps, if we are given any group, we can pick any one of its subgroups and from it get a transitive action of the group on some space. Any two conjugate subgroups will lead to isomorphic actions by this construction, and so the distinct transitive actions of a group correspond exactly to its subgroup conjugacy classes. One can think of the multiplication table that defines a group algebraically as a system of equations between monomials of the group elements. A group action is a  solution to the system of monomial equations, a concrete permutational representation of their algebraic structure, and finding the subgroup conjugacy classes is the method of solution. For finite groups the method is algorithmic \cite{GAP4}. It is also much aided by the Classification of Finite Simple Groups (CFSG). Finding all the subgroups of an infinite group however is generally not straightforward, and so for infinite groups the conjugacy class approach is mostly implicit. Some infinite groups of broad importance have been partly analysed in this manner however \cite{brazil1994maximal,macpherson1990subgroups,macpherson1991large,ball1966maximal,rosenberg1958structure,macpherson1992maximal,antoneli2006maximal}. 

We shall use basic concepts from First Order Logic and Model Theory. In particular we shall use the notions of definable closure and definably closed sets, see \cite{hodges1993model,marker2006model} for comprehensive treatments. For the emergence of FOL as the common language for essentially all mathematical theories see \cite{sep-logic-firstorder-emergence} and references therein. For a lucid exposition of the basics of its relation to set theory see \cite{hanskamp}. We shall also discuss the projection lattices of von Neumann algebras in the context of quantum fields. A thorough reference on von Neumann algebras is \cite{kadison1997fundamentals}. For a breathtaking rapid review of this deep subject see \cite{jonesrapid}. Finally, the subject of Quantum Logic itself has been massively researched, and any attempt at summarising the literature here is futile. An excellent review is given in \cite{sep-qt-quantlog}, together with a list of core references.

\section{The interpretation of meets and joins in FOL}

Let us see first how the notion of a propositional lattice enters into theories of Physics. A theory of a physical system at the kinematical level\footnote{By this we mean statements descriptive of single observations of an individual system, thus excluding dynamics, which necessarily refer to relations between multiple observations. These are usually expressed in terms of ranges of values of physical magnitudes e.g. ``the energy of the system is between 125.2 and 125.5 GeV''.} consists of a pair $(S, T)$, where $S$ is the set of possible statements about the system and $T$ is a set of truth assignments on $S$ interpreting experimental outcomes. Every observation of the system assigns True to a subset of $S$. The set of truth assignments $T$ defines a natural implication structure on the set of statements $S$. If every truth assignment $t \inn T$ that assigns True to a statement $p$, also assigns True to a statement $q$, we say that $p$ implies $q$. If also $q$ implies $p$ they are implicationally equivalent $q \sim p$ and the set $S$ can be split into equivalence classes of such statements. The set of equivalence classes $S / \! \! \sim$ is a poset $\mathcal{P}$ with the partial order relation being the implication defined by $T$ as above. Observe now that for the set of truth assignments $T$ to merit its name, it must be closed under finite meets, i.e. the theory $(S, T)$ must be finitely consistent \cite{turner2021ultrafilters}. Finite consistency means that the conjunction of any finite set of True statements is also True, and thus also an element of $\mathcal{P}$. In Physics, where statements are rendered True or False by observations, we can reasonably extend the requirement of consistency to arbitrary subsets of $\mathcal{P}$. This makes $\mathcal{P}$ into a complete meet-semilattice, which is always a complete lattice $\mathcal{L}$ \cite{birkhoff1940lattice,hawaii}. This line of argument is essentially the same as that of the Geneva school of quantum logic (see \cite{piron1976foundations} esp. Thm. 2.1, and \cite{moore2009operational} Sec. 3).

Let $\mathcal{L}$ be a complete lattice and $(G, \mathcal{L})$ the action of its group of automorphisms. The action $(G, \mathcal{L})$ is in general a subdirect product of actions $\mathbf{Sub}(\prod_{I}(G_{I}, \mathcal{L}_{I}))$ where $\{\mathcal{L}_{I}\}$ is a partition of $\mathcal{L}$. Assume now that the following injectivity condition, which we shall refer to as condition (S), holds for the automorphisms of $\mathcal{L}$: 
\begin{equation}
\forall \, l, l' \inn \mathcal{L}, \;\; l \neq l' \implies G_{l_{\downarrow}} \neq G_{l'_{\downarrow}}
\end{equation}
Assume moreover that $\mathcal{L}$ is atomistic and let $A \subset \mathcal{L}$ be its set of atoms. Define the function: 
\begin{equation}
    \mathbf{a}: \mathcal{L} \rightarrow \mathscr{P}(A), \;\; \mathbf{a}(l) = \{a \; | \; a \leq l \}
\end{equation}
Since $\mathcal{L}$ is atomistic, $l = \vee \mathbf{a}(l)$ and because of the uniqueness of the join $\mathbf{a}$ is injective, so that $G_{\mathbf{a}(l)}$ fully induces $G_{l_{\downarrow}}$. Under condition (S):
\begin{equation}
    l \neq l' \implies G_{\mathbf{a}(l)} \neq G_{\mathbf{a}(l')}
\end{equation}
We have then a 1-to-1 map
\begin{equation}
    \mathbf{f}  : \mathcal{L} \rightarrow \mathcal{L}_{\mathrm{fix}}, \;\; \mathbf{f}(l) = \mathbf{fix}(G_{\mathbf{a}(l)})
\end{equation}
We see that $\mathbf{f}$ maps the elements of the lattice to fixsets of the action $(G, A)$, which is induced by $(G, \mathcal{L})$ on the set $A$ of atomic elements of $\mathcal{L}$. But the atomicity of $\mathcal{L}$ means that $(G, A)$ in turn induces $(G, \mathcal{L})$, because elements of $\mathcal{L}$ correspond uniquely to elements of $\mathscr{P}(A)$. We can then restrict our attention to $(G, A)$. If $\mathcal{L}$ is assumed atomic but not atomistic, this is no longer true, because the joins contain extra information on top of the atoms contained in their lower cone. Such a situation, in which to determine $(G, \mathcal{L})$ one would need to know the action on the extra variables that determine the joins, but are not lattice elements themselves, does not arise in the lattices of known physical theories and we shall not examine it further. The image of $\mathbf{f}$ is always a closure operation on $\mathcal{L}_{\mathrm{fix}}$, 
\begin{equation}
   \mathbf{cl}: \mathcal{L}_{\mathrm{fix}} \rightarrow \mathcal{L}_{\mathrm{fix}}, \;\; \mathbf{cl}(\Delta) = \wedge\{\mathbf{f}(l) \;|\; \Delta \subseteq \mathbf{f}(l)\} 
\end{equation}
 With one exception, which we shall examine later in this section, in all cases of interest in Physics the closure operation is trivial, $\mathbf{f}$ is bijective and we have a lattice isomorphism $\mathcal{L} \cong \mathcal{L}_{\mathrm{fix}}$. When this is the case, the meets and joins of $\mathcal{L}$ can be expressed directly in terms of the action $(G, A)$:
\begin{subequations}
\begin{align}
& \mathbf{a}(l \wedge l') = \mathbf{fix}(\langle G_{\mathbf{a}(l)}, G_{\mathbf{a}(l')} \rangle) =  \mathbf{fix}(G_{\mathbf{a}(l) \cap \mathbf{a}(l')})  \\ 
& \mathbf{a}(l \vee l') = \mathbf{fix}(G_{\mathbf{a}(l)} \cap G_{\mathbf{a}(l')}) =  \mathbf{fix}(G_{\mathbf{a}(l) \cup \mathbf{a}(l')}) 
\end{align}
\end{subequations}
These expressions generalise to arbitrary collections of lattice elements, by the completeness of the pointwise stabiliser lattice $\mathcal{L}_{\mathrm{stab}}$:
\begin{subequations}
\begin{align}
& \mathbf{a}(\wedge \, l_I) = \mathbf{fix}(\langle G_{\mathbf{a}(l_I)}\rangle) =  \mathbf{fix}(G_{\cap \mathbf{a}(l_I)}) = \cap \mathbf{a}(l_{I})  \\ 
& \mathbf{a} \, (\vee \, l_I) = \mathbf{fix}(\cap G_{\mathbf{a}(l_I)}) =  \mathbf{fix}(G_{\cup \mathbf{a}(l_I)}) \supseteq \cup \mathbf{a}(l_{I})
\end{align}
\end{subequations}
How can we interpret these formulas? The action $(G, A)$ defines a relational structure $\mathcal{R}$ in FOL with domain $A$. The orbits of the action on $A^n, n \inn \mathbb{N}$ are the $n$-ary relations and the fixsets of the automorphism group action correspond bijectively to the definably closed sets of the structure. From the relational perspective, these are the subsets whose elements can be defined using their relations to other elements of the initial set only. Starting from any subset $X \subseteq A$, one can look for points in $A \setminus X$ that can be specified by points in $X$. By the inclusion of all these points one gets the definable closure $\mathbf{dcl}(X)$, allowing us to express meets and joins in the most concise form:
\begin{subequations}
\begin{align}
& \mathbf{a}(\wedge \, l_I) = \mathbf{dcl}(\cap \mathbf{a}(l_I)) \\
& \mathbf{a}(\vee \, l_I) = \mathbf{dcl}(\cup \mathbf{a}(l_I))
\end{align}
\end{subequations}

A standard example is that of a projective space. The points of $A$ are the rays, and the defining relations are ternary and up, because $PGL$ is 2-transitive in its action on the rays. The ternary relations consist of triplets of coplanar rays. Every plane of the projective space corresponds to one such ternary relation. Consider two rays $\psi_1$, $\psi_2$, and let $P$ be the plane they define. The definable closure of their union $\mathbf{dcl}(\{\psi_1, \psi_2\})$ will include all the solutions to the ``equation'' $(\psi_1$, $\psi_2, x)$, i.e all rays $x$ coplanar with $\psi_1$, $\psi_2$. These are of course the set $\{\psi \; | \; \psi \inn P \}$, i.e. the plane defined by the original pair of rays. Adding rays to the original set will bring higher arity relations into play. For example, quaternary relations are defined by pairs of pairs of points with a given harmonic ratio, and so on. More generally, either from the orbits of the action of $PGL$ on the rays or directly from the standard defining relations of the projective space over a complex vector space, one can see that $\mathbf{dcl}(\{\psi_I\}) = \mathbf{span}(\{\psi_I\})$, i.e. the definable closure of any collection of rays will be the projective subspace they span. Their lattice, $\mathcal{L}_{\mathrm{dcl}}$, is isomorphic to the lattice of closed linear subspaces of the corresponding Hilbert space, which is the original Quantum Logic of \cite{logic_of_qm}. We should note that although Birkhoff and von Neumann derive $\mathcal{L}$ by finding the preimage in Hilbert space of statements about the values of observables in a quantum system, they remark that the algebraic structure of $\mathcal{L}$ does not use the topological completion involved in the construction of a Hilbert space (\cite{logic_of_qm} sec. 12). The role of this completion is to support a (countably additive) measure of probability, which is not needed for the algebraic implication structure expressed in $\mathcal{L}$. From a model-theoretic perspective, there is a difference between the projectivization of an infinite-dimensional complex vector space $P(V)$, and its corresponding projective Hilbert space $P(H)$. While $P(V)$ can be formulated in standard finitary FOL ($L_{\omega\omega}$, see \cite{sep-logic-infinitary} for definitions and references), $P(H)$ requires passing to its infinitary extension known as Continuous First Order logic (\cite{yaacov2008model} ch. 15, \cite{goldbring2012definable}),  a fragment of $L_{\omega_{1}\omega}$ (\cite{yaacov2017metric} sec. 10). As shown in \cite{yaacov2008model,goldbring2012definable}, the definably closed sets of the Hilbert space infinitary first order structure are its closed linear subspaces, confirming that there is an isomorphism between $\mathcal{L}_{\mathrm{dcl}}$ in the finitary first order structure $P(V)$ and its infinitary counterpart in $P(H)$. See also  sec. 3 in \cite{holland1995orthomodularity}, for a different but related perspective on the isomorphism between $P(V)$ and $P(H)$.

The case of the two-dimensional Hilbert space, whose algebra of bounded operators is the von Neumann factor $I_{2}$, describing what is called a qubit, is unique in being the only example among known physical theories of a non-trivial closure on the fixset lattice of its namespace. The fixset lattice of $PGL(2, \mathbb{C})$ consists, bottom to top, of the empty set, the layer of atomic names, and the whole space. This is because a join of any two rays in two-dimensional projective space is the space itself. The automorphism group of this lattice is a subbdirect product of $Sym(2^{\aleph_0})$ acting on the layer of atoms $A$, and the trivial action on the whole space and the empty set. The corresponding namespace is an uncountable pure set and its fixset lattice is the powerset $\mathscr{P}(A)$. The physical lattice is a closure on $\mathscr{P}(A)$, in which every super-atomic fixset is mapped to the whole space. One notices that the enhanced symmetry of $\mathcal{L}_{\mathrm{fix}}$ actually is the naming group of a classical particle system, as we shall see in the next paragraph. This would seem to suggest that a single qubit has an underlying classical model. For a free qubit such a classical hidden variable model has been explicitly constructed in \cite{kochen1967problem}. It is not clear how the enhanced symmetry of $\mathcal{L}_{\mathrm{fix}}$ is related to the possibility of introducing a hidden variable theory in this example (see remark 3).

For the lattice of classical particle mechanics, i.e. classical systems with a finite number of degrees of freedom, two options are put forth in \cite{logic_of_qm}. The first option allows for exact measurements of the state of the system, down to the exact position and momentum of every particle\footnote{We could call this the Laplacean theory of measurement, after the great mathematician's infamous encounter with Napoleon.}. These are called ``maximal observations'' in \cite{logic_of_qm} and they are the atoms of $\mathcal{L}$. In this case the naming group is $Sym(\Phi)$, the namespace that it defines is the classical phase space $\Phi$ viewed as an unstructured set with no relations\footnote{Or equivalently all trivial relations of every arity, i.e. the relations that contain all possible tuples of that arity if viewed as the canonical structure defined by the orbits of $Sym(\Phi)$ on $\Phi^{n}$.}. $\mathcal{L}$ is the fixset lattice of $Sym(\Phi)$ which is $\mathscr{P}(\Phi)$, the full powerset lattice of $\Phi$. This view is ultimately rejected by the authors of \cite{logic_of_qm} in favour of a second option, the non-atomic lattice of Lebesgue measurable mod Lebesgue-null subsets of the phase space $\mathcal{L}_{\mathrm{leb}}$. It is constructed by restricting first to the sublattice of Lebesque measurable sets in $\mathscr{P}(\Phi)$ and then taking equivalent classes by modding out sets of null measure. The join and meet operations are still set union and intersection by choosing any representative from the classes. The authors justify their choice by the impossibility of maximal observations in actual measurements. At the same time it allows for the introduction of probabilistic calculus and rigorous ergodic theory in classical mechanics. When $\Phi$ is compact, the measures of these sets properly normalised can be interpreted as the probability of the system being in an atomic state contained in the set, which itself can not individually be measured, or can sometimes be used to define conditional ``transition probabilities'' between such inexact measurements through the measure of the overlaps of the corresponding sets. In contrast, the full powerset lattice per se does not allow the introduction of probability either as measure of sets or as transition probabilities between atomic states. In the latter case the continuity of independent atomic pure states in $\Phi$ will not allow probabilities to be normalised.

To progress from the description of particle systems to theories of fields, i.e. systems with infinite degrees of freedom that extend in space and time, one needs to incorporate additional constraints coming from the continuity and causal structure of spacetime. We shall not discuss the constraints of microcausality. Instead, we shall limit ourselves to the description of the local physics of an observer. For classical field theory, the naming group allowing for maximal observations would be $Sym(2^{2^{\aleph_0}})$. Imposing continuity on field configurations allows us to define them from their values on a countable dense subset of spacetime. If, in addition, we follow \cite{logic_of_qm} in rejecting the possibility of maximal observations and instead limit possible outcomes to measurable sets of field values at each point, we can introduce probability measures in the space of field configurations \cite{bar2011wiener}. The lattice of Wiener measurable sets of field configurations $\mathcal{L}_{\mathrm{w}}$ plays a role directly analogous to $\mathcal{L}_{\mathrm{leb}}$ in classical particle theories. For both particles and fields Hamiltonian dynamics preserve the measure of the lattice elements and the corresponding phase spaces are symplectic manifolds, which for field theories such as Electromagnetism and General Relativity are infinite dimensional \cite{marsden1974reduction}. Quantum fields are described by a net of local operator algebras \cite{haag2012local} which is a map from open double cones in spacetime to operator algebras on a separable Hilbert space for each double cone. For a long time it was deemed that these local algebras have to be the hyperfinite type $III_{1}$ von Neumann factor (\cite{halvorson2006algebraic}, sec. 2.5.3). However, recent research suggests that the (necessary) inclusion of an observer in closed universes such as de Sitter universes, which our own is often assumed to resemble, leads to the algebras of local observables being the hyperfinite factor of type $II_{1}$, rather than $III_{1}$ \cite{chandrasekaran2023algebra,witten2023algebras}. It is interesting to note that while type $II_{1}$ and $III_{1}$ factor representations are constructed by infinite tensor products of finite dimensional Hilbert spaces through the GNS representation \cite{witten2022does}, the type $II_{1}$ hyperfinite factor (call it $R$) also admits a decomposition into an infinite tensor product of copies of itself, $R=\overline{\otimes}_{n \in \mathbb{N}}(R, \tau)$, taking the trace $\tau$ of each copy as the preferred state for constructing the GNS representation \cite{isono2019unique,mcduff1970central}. This suggests that a construction analogous to the one for classical field theory may apply. In particular, the projection lattice of $R$ is the quantum equivalent of the lattice $\mathcal{L}_{\mathrm{leb}}$ of non-maximal observations in classical particle mechanics, because it consists of non-atomic projections whose finite rank can be interpreted as a probability measure \cite{von1981continuous,redei1996john}. If we follow the same steps as in the classical case, attaching a factor $R$ to each of the points of a countable dense subset of a local region, representing non-maximal measurements of the field operators, we are led to a single copy of $R$ representing measurements on the whole region through the inverse of the decomposition mentioned above. The choice of subset does not matter as homeomorphism and diffeomorphism groups acting on connected manifolds are transitive on countable dense subsets \cite{ford1954homeomorphism}.

In field theory, both classical and quantum, as well as in classical particle mechanics, the lattices labeling experimental outcomes are non-atomic sublattices of the corresponding atomistic fixset lattices of the Symmetric or Projective Linear group acting on the set of atoms. In quantum field theory the projection lattices of type $II$-$III$ factors are complete sublattices of the fixset lattice of $PGL(\aleph_0, \mathbb{C})$. Their elements can be mapped to moietous subspaces\footnote{A subspace of $P(H)$ is called moietous when both its dimension and codimension are infinite. The non-atomic projection lattices of von Neumann factors are all embedded in the sublattice of mutually moietous projections, i.e. projections to moietous subspaces whose meet and join are always also moietous.} in the fixset lattice of $PGL$. It is natural to ask, given a non-atomic physical lattice, whether the atomic background can be detected by examining the action of its automorphism group. Clearly, condition (S) still applies to these lattices. However the construction given above can not proceed because there is no layer of atoms inducing the action on the whole lattice. A well known theorem, applied to the classical case, gives us a hint on how the question could be approached. Take the lattice $\mathcal{L}_{\mathrm{leb}}$ describing experimental outcomes in classical particle systems. Stone's representation theorem tells us that this lattice, being distributive, can be represented as an algebra of sets of ultrafilters that contain a given element of the lattice. Let $U$ be the set of ultrafilters of $\mathcal{L}_{\mathrm{leb}}$. Every $l \inn \mathcal{L}_{leb}$ belongs to at least one ultrafilter since its own upper cone can be extended to an ultrafilter by Zorn's lemma. We have then a map:
\begin{equation}
    \mathbf{u}: \mathcal{L}_{leb} \rightarrow \mathscr{P}(U), \;\; \mathbf{u}(l)= \{u \; | \; l \inn u \}
\end{equation}
Under the assumptions of distibutivity and complementation, which $\mathcal{L}_{\mathrm{leb}}$ satisfies, $\mathbf{u}$ is injective and it provides an algebra of sets representation of $\mathcal{L}_{\mathrm{leb}}$. At the same time there are in $U$ ultrafilters that converge to each point of the original point set $A$ from which $\mathcal{L}_{\mathrm{leb}}$ is constructed, in the sense of containing every open neighborhood of a point in $A$. In fact, there are at least $\aleph_0$ of them for each of the points in the original point set $A$ (\cite{halvorson2001nature}, prop. 3). Therefore, in the action induced by $(G, \mathcal{L}_{leb})$ on $\mathscr{P}(U)$ there will be an orbit $O_{con}$ consisting of the sets: $u_p = \{u \inn U \; | \; u \; \mathrm{converges \; to} \; p\}$ for all $p \inn A$. The action induced by $(G, \mathcal{L}_{leb})$ on $O_{con}$ will be isomorphic and completely `locked in' to $(Borel(A), A)$, through the bijection $p \leftrightarrow u_p$. $Borel(A)$ is the highly transitive subgroup of $Sym(A)$ that preserves measurability. Its topological closure is $(Sym(A), A)$, the atomic action whose fixset lattice naturally embeds $\mathcal{L}_{\mathrm{leb}}$. And since it is $(Borel(A), A)$ that ultimately determines the action on $\mathscr{P}(U)$, one observes that the whole subdirect product action on $\mathscr{P}(U)$ would be seen to be induced by its $O_{con}$ orbit. Note that the elements of $O_{con}$ are not in $Img(\mathbf{u})$, since the atoms themselves are not elements of $\mathcal{L}_{\mathrm{leb}}$ (they are of measure zero, and so all mapped to the equivalence class of the empty set in $\mathcal{L}_{\mathrm{leb}}$). On its face, this way of recovering the atomic substrate of $\mathcal{L}_{\mathrm{leb}}$, which could then be said to be `crypto-atomic', is of little or no practical value since the action $(G, \mathcal{L}_{leb})$ is not known explicitly in algebraic terms. It is nonetheless interesting to see how the conditions of injectivity of $\mathbf{u}$ and existence of convergent ultrafilters, together amount to a sort of `crypto-atomicism' for a non-atomic lattice, where certain sets of its ultrafilters (the ones containing ultrafilters designated in hindsight as convergent on a point) are seen to play the role of atoms. To our knowledge the question of convergent ultrafilters in the projection lattices of von Neumann algebras has not been explicitly addressed. 


We conclude this section with some observations. We have shown that atomistic physical proposition lattices are produced by relational structures in FOL, whose domains correspond to the maximal observations of the theory. Maximal observations are the minimal subsystems, and the requirement that every subsystem can be independently labeled permits us to analyse systems down to their atomic layer. The relations between the atoms determine the full lattice through the operation of definable closure. Thus the combined requirements of atomicism and condition (S) amount to total reductionism to the atomic layer of the lattice: Condition (S) allows us to take sections, isolating smaller and smaller subsystems (i.e. taking increasingly precise measurements), while with atomicism we postulate first that the implication chains terminate and secondly that all other lattice labels are simple sets of the terminal nodes, resulting in a bottom-up, upwards determinacy for the lattice. Combining the upwards and downwards directions we have encoded all the lattice information in the structure of its atomic layer. Non-atomic physical lattices satisfy condition (S), but lacking the upwards determinacy of atomicism they are not reductionist in this sense. They are however embeddable in the atomic lattices in a natural way. One may notice that the concept of probability, central to Quantum Mechanics, does not at all enter into these considerations. This comes from the fact that the lattice captures the implication structure of the theory, expressing only relations of certainty. Hence, in order to arrive at the formulation of Quantum Mechanics, one needs to ask which relational structures in FOL can model atomic transition probabilities. As it turns out, one can narrow the search dramatically by taking first into account the manner in which the naming group allows a system to be analysed into its subsystems - specifically by demanding that this process is always unambiguous when the labels in a subsystem are held fixed. This will be the subject of the next section.

\section{Jordan groups as naming groups of physical theories}

In Physics every experimental observation refers to a part of the universe. Therefore the notion of a physical system is inseparable from that of being a subsystem. At the kinematical level of the lattice of implication $\mathcal{L}$, the natural definition of a subsystem is: a bounded subset of lattice elements that is implicationally closed. A subset $S_{sub} \subseteq \mathcal{L}$ is implicationally closed if:
\begin{equation}\label{impl}
\forall s \inn S_{sub},\;l \inn \mathcal{L}, \;\; l \leq s \implies l \inn S_{sub}
\end{equation}
Boundedness (from above, as $\mathbf{0} \inn S_{sub}$ by the requirement of implicational closure) is needed to guarantee the unicity of the subsystem. The maximum element can be thought of as corresponding to the statement ``the subsystem exists''. We can read \eqref{impl} as saying that $S_{sub}$ contains the lower cones of all its elements. If $s_{max}$ is the maximum element, meaning that it is implied by every $s \inn S_{sub}$, then $S_{sub} = s_{max\downarrow}$. Hence, at the kinematical level of description a subsystem is a lower cone of $\mathcal{L}$. This is not the usual definition of a subsystem, but a more general one that includes the standard one as a special case. Usually one defines a subsystem as a material fragment of the original system. The maximal observations corresponding to this fragment considered by itself can also be found\footnote{The statements of the subsystem only refer to the subsystem while those of the larger system refer to phenomena that include the subsystem. We expect however to find the original maximal observations as parts of statements describing maximal observations in the bigger system.} among the maximal observations of the original system. We can postulate the preservation of maximal observations because otherwise extending a system would interfere with the theoretical measurement accuracy in the original system, which is something we do not expect. Looking at a subsystem as a system in itself it follows that one should be able to fix the labels in any subsystem, i.e. lower cone, without labeling anything else outside the subsystem. This is the content of condition (S). Mathematically it is formulated as a requirement on the action $(G, \mathcal{L})$ of the group of automorphisms of $\mathcal{L}$. The allowed relabelings after fixing the labels in a lower cone $l_{\downarrow}$ form the group $G_{l_{\downarrow}}$, and condition (S) declares that $\mathbf{fix}(G_{l_{\downarrow}}) = l_{\downarrow} \; \forall l \inn \mathcal{L}$. We saw the consequences of this in the previous section. But there is more to be said about subsystem extensions, leading to a condition significantly stronger than (S).

When $\mathcal{L}$ is atomistic, condition (S) implies that every set of atoms corresponding to a lattice element is a fixset of the action $(G, A)$ on the atomic layer. Assuming that there are no permanent properties, i.e. unary relations, distinguishing maximal observations, the action $(G, A)$ is transitive, which means that the label of a single maximal observation is arbitrary. An example of such a unary relation is the case of quantum numbers distinguishing superselection sectors in quantum mechanics. In this case the action on the atomic layer is a direct product of transitive actions on each sector. We want to examine how subsystems within each transitive sector (i.e. with the same quantum numbers) behave under extension. The pointwise stabiliser $G_{\mathbf{a}(l)}, l \inn \mathcal{L}$ acts on the complement $A \setminus \mathbf{a}(l)$ and in general will partition $A \setminus \mathbf{a}(l)$ into a set of orbits. Consider now $l_{\downarrow}$ as a subsystem of $\mathcal{L}$, the latter being the lattice of a transitive component. Let $a \inn A$ be a maximal observation outside $\mathbf{a}(l)$. If there are more than one orbits when extending the system from $l_{\downarrow}$ to $\mathcal{L}$, extensions into which $a$ falls into different orbits would be inequivalent. Because, since $(G, A)$ is transitive, assigning one of the remaining labels to $a$ from a different orbit would require changing the labels in $\mathbf{a}(l)$. This would make it impossible to extend a subsystem to a supersystem while retaining the labels already assigned. If one insisted in holding on to the labels already assigned to $\mathbf{a}(l)$, then extensions starting from different orbits would not be connected by an automorphisms although their unary relations (quantum numbers) were identical. We are therefore led to postulate the transitivity of $G_{\mathbf{a}(l)}$ on $A \setminus \mathbf{a}(l)$ for all $l \inn \mathcal{L}$. We call this condition (T). An immediate consequence is that $(G, A)$ is not only transitive but also primitive. In an imprimitive action the blocks of imprimitivity are permuted amongst themselves. By condition (S) an atom $a \inn A$ can be individually stabilised, but condition (T) requires that $G_{a}$ be transitive in $A \setminus \{a\}$. This is impossible if $a$ belongs to a block of imprimitivity, as its pointwise stabilisation would stabilise the block setwise. 

The property of being transitive on the complement of a fixset is known in group theory as the Jordan property. Condition (T) can be then formulated as the requirement that all fixsets of $(G, A)$ have the Jordan property. Returning to the original question of which permutation groups are appropriate as naming groups of physical theories, we see that we have now a possible characterization: They are the closed primitive permutation groups whose every fixset has the Jordan property. In a $k$-transitive action any fixset of cardinality up to $k-1$ has the Jordan property. A fixset of cardinality greater or equal to $k$ that has the Jordan property is called\footnote{In the literature a complementary terminology is used. The standard term is ``Jordan set'', referring to the points that are \emph{not} fixed. Here we are interested mainly in the fixsets themselves, so it is easier to talk about ``Jordan fixsets'' than ``the complements of Jordan sets''.} a proper Jordan fixset. A permutation group admitting a proper Jordan fixset is called a Jordan group. Highly transitive Jordan groups\footnote{Known examples include among others the homeomorphism and diffeomorphism groups of manifolds, but there is no  classification \cite{bhattacharjee2006notes}.} are not topologically closed, except for the Symmetric Group itself, which is the closure of every highly transitive group. It is the naming group of classical mechanics, and because of its continuum of independent states it cannot incorporate atomic transition probabilities. So although it is admissible as a naming group, it cannot express the randomness of quantum phenomena. The naming groups we are after are therefore a subclass of non-highly-transitive primitive Jordan groups. Rather remarkably, considering the generality in their definition, primitive Jordan groups that are not highly transitive have been classified.

The classification of non-highly-transitive infinite primitive Jordan groups is a major result in Group Theory and Infinite Combinatorics. Finite Jordan groups were first studied by Camille Jordan \cite{jordan1871theoremes,neumann2006concept} and were fully classified following CFSG. All of them preserve finite geometric structures \cite{kantor1985homogeneous,neumann1985some,cherlin1985N0} and are at least $2$-transitive. In the infinite case, a new class of non-geometric Jordan groups appears. By the classification theorem, these are found to preserve linear or semilinear orders and other structures derived from these, taking advantage of Hilbert Hotel moves in infinite linear orders to preserve transitivity in a way that is impossible for finite structures. This class of structures does not satisfy condition (T). In any of the structures in this class, preserved by a Jordan group of transitivity $k$, fixing a generic set of $k$ points will result in a fixset not transitive on its complement as can be easily seen by inspection of the diagrams and definitions of the structures \cite{bhattacharjee2006notes}. Proper Jordan fixsets are the exception in the fixset lattices of these group actions, rather than the rule. Within the class of non-geometric structures with Jordan automorphism groups one can distinguish two subclasses:
\begin{enumerate}[label=\roman*), ref=\roman*,  topsep=0pt]
\item The first subclass contains dense linear orders and other weaker relations derived from them. Those can be of three types, called linear betweenness relations, circular orders and circular separation relations. The automorphism groups of these structures are distinguished by being highly homogeneous but not highly transitive and they were the first infinite Jordan groups to be classified, by Cameron \cite{cameron1976transitivity}.   
\item The second subclass contains structures from the family of semilinear orders (also called trees) and relations defined from these called B (or general betweenness), C and D relations. They were classified by Adeleke and Neumann in \cite{adeleke1996primitive}. C relations are defined on sets of maximal chains of a semilinear order while B and D relations are produced by semilinear orders and C relations by forgetting the direction of the semilinear order in much the same way as linear betweenness relations result from symmetrising a linear order \cite{adeleke1998relations}. It is also possible to define limit structures of B and D relations with Jordan automorphism groups \cite{adeleke2013irregular,bhattacharjee2006jordan,almazaydeh2022jordan,bradley2023limits,almazaydeh2019infinite}. 
\end{enumerate}
Since the structures above do not satisfy condition (T), they cannot be used as naming groups for physical lattices. It is interesting however to note that the grounding structures of the two subclasses, linear and semilinear orders, are in fact used in Physics, but to coordinatise time in physical processes. Open-ended dense linear orders can represent time for most physical systems. Dense semilinear orders are used for the more exotic model of time in the many-worlds interpretation of quantum mechanics.

Geometric Jordan groups, of which all known examples satisfy condition (T), were classified\footnote{This was achieved after dropping the assumption of primitivity of Jordan sets, i.e. the assumption that the action of the pointwise stabiliser on the complement of some Jordan fixset is primitive. Limits of B and D relations also appeared in \cite{adeleke1996classification}, as they only have non-primitive Jordan sets.} by Adeleke and Macpherson in \cite{adeleke1996classification}, thereby completing the classification of primitive Jordan groups. As in the finite case, infinite geometric Jordan groups are so called because they preserve Steiner systems, which are geometric structures of points organised in blocks of equal cardinality. These will be our main focus, selected because of their compliance with condition (T). Steiner systems are found in the classification also in the form of limits, or chains: in infinite cardinalities a type of quasi-geometric Jordan groups appears which, like the non-geometric examples, have no analogue in the finite case. They preserve infinite towers of Steiner systems nested inside so-called ``peculiar'' C relations \cite{adeleke1995semilinear,johnson2002constructions}. The automorphism groups of each of the Steiner systems making up the tower also satisfy condition (T), although the Jordan group preserving the limit does not. Hence we characterise them as quasi-geometric, because although they themselves are not geometric, they contain infinite chains of geometric Jordan setwise stabilisers (the groups preserving the individual Steiner systems). We have no reason to make an exception to condition (T) for these quasi-geometric groups, but the fact that they contain Steiner systems, as well as their connection to C relations (see remark 1) could mean that there is an application to Physics. We shall for now stick with Steiner systems themselves as candidate structures for a physical namespace.

A Steiner $k$-system, $k \geq 2$, is a set $\Omega$ of points and blocks $B_I \subset \Omega$ of equal cardinality, where every set of $k$ points is contained in exactly one of the $B_I$. For real physical systems $\Omega$ will be uncountable. From a Steiner $k$-system $k \geq 3$ one can construct a Steiner $(k - 1)$-system by a process called derivation. The stabiliser $G_{a}$ preserves a Steiner $(k - 1)$-system on $\Omega \setminus \{a\}$ with blocks $B_{J} \setminus \{a\}$ where $J \! \subset \! I$ indexes the blocks of the $k$-system that contain $a$. The inverse process is called extension: Adding a point $a$ to $\Omega$ and constructing a Steiner $(k+1)$-system on $\Omega \cup \{a\}$ such that the derivation of the new system at $a$ is the original system. All infinite Steiner systems can be extended \cite{beutelspacher1994transfinite}. Moreover, when the Steiner system has a Jordan automorphism group its derivations and extensions have Jordan automorphism groups (\cite{adeleke1996classification} lem. 2.2.6 \& 2.2.7). Thus the class of infinite Steiner systems with Jordan automorphism groups is closed under derivation and extension and can be reduced to the class of Jordan Steiner $2$-systems. Additionally we can observe that a Steiner $k$-system, $k \geq 3$, all sets of cardinality up to $k - 1$ are improper Jordan fixsets. This means that the lattice of fixsets of the automorhism group will look like a boolean lattice up to cardinality $k-1$ but switch to non-boolean in cardinality $k$, where whole blocks\footnote{It is unknown if in all Jordan Steiner $k$-systems the pointwise stabilisation of $k$ points stabilises the whole block pointwise. By definition, the block is stabilised setwise. But in all known examples of Jordan Steiner systems blocks are stabilized pointwise.} are stabilized. So in addition to not containing any new information compared to their `base' Steiner $2$-systems, they distinguish a natural number ($k-2$, the number of extensions), which from a physical point of view is undesirable. We can then limit our attention to Steiner $2$-systems.

The only known Steiner $2$-systems with Jordan automorphism groups in cardinality $2^{\aleph_0}$ are affine and projective spaces. In an affine space the blocks are the lines, defined uniquely by pairs of points. In a projective space the points are the rays of the corresponding vector space. Every pair of rays defines a unique two dimensional subspace which is the Steiner block. Between $AFL$ and $PGL$, it is an easy choice for a theory with atomic transition probabilities for any pair of atoms. Because the affine group is the wreath product of the General Linear group of $V$ with the translations by vectors in $V$, $AFL(V) = GL(V) \wr V$, when $V$ is infinite the translations will make it impossible to normalise any such probability measure, as is the case also in classical mechanics. But it is not known if these are the only examples. In finite cardinalities all Steiner systems are connected to linear spaces (see the references in \cite{adeleke1996infinite} rem. 5.8.4). In countably infinite cardinality, examples with Jordan automorphism groups have been constructed for all $k$ \cite{hrushovski1993new}. But in uncountable cardinalities little is known \cite{cameron2002infinite}. There are then two possibilities. If more Jordan Steiner $2$-systems are discovered in uncountable cardinalities, we would need an extension of Gleason's theorem \cite{gleason1957measures} to decide in which of these structures can support transition probabilities, and for which of them this is actually the only option, as Gleason showed for the Born rule in Hilbert space. Indications that Gleason's theorem is essentially combinatorial in nature are given in \cite{pitowsky1998infinite}. If on the other hand it could be proved that these are the only options then all that would be left to decide would be the field over which the projective space is defined. For this we feel that there exist sufficient arguments to show that it has to be the complex numbers (see \cite{aaronson2004quantum, kapustin2013there}). Among these a rather striking one is that of Wootters \cite{wootters1981statistical}, who shows that random selection of rays in Hilbert space results in a uniform distribution of mutual transition probabilities on the probability simplex only if the field is $\mathbb{C}$.

There have been two main approaches to quantum logic in the vast literature that followed \cite{logic_of_qm}. The first, introduced by Mackey in \cite{mackey1963mathematical}, starts from atomic transition probabilities as the fundamental concept but fails to derive the structure of Hilbert space, which then has to be postulated. The other, pioneered by Jauch and Piron, uses lattice theoretic methods and culminated in Piron's theorem \cite{piron1964axiomatique}, which in combination with Solèr's theorem \cite{soler1995characterization} does actually recover the lattice of subspaces of the Hilbert space but at the cost of one not easily interpretable axiom, the ``covering law''. Here we have seen that one can apply strong selection to candidate theories expressed by condition (T), using very fundamental notions that precede the introduction of probability and dynamics in the description. Moreover, all the structure in our approach is contained in the atomic layer, and we do not make additional assumptions at the level of the lattice such as orthomodularity or complementation. It does however hinge on the classification of infinite Jordan Steiner $2$-systems, which is an open problem.

\section{Remarks}

1. One obvious absence from the above discussion is path integrals. Although not easy to define rigorously, they play a central role in quantum physics and especially in quantum field theory. Hilbert spaces can be formally defined by path integrals but in the combinatorial/group theoretic framework it is not clear how to formulate the connection. There is a seemingly superficial similarity with the somewhat mysterious limits of Steiner systems with quasi-geometric Jordan automorphism groups, but it could be more. These Steiner systems live inside ``peculiar'' C relations. The maximal chains in a semilinear order that satisfy the C relation resemble piecewise linear paths used in the definition of path integral and at least in one case an intriguing connection to vector spaces exists (see \cite{adeleke1996classification} example below hypothesis 5.1.1). But we have no indication of whether Hilbert spaces or projective spaces can be constructed inside C relations.

\noindent 2. Connections between Physics and Model Theory have been pointed out before, by Zilber and collaborators \cite{zilber2014perfect,cruz2015geometric,cruz2021around}, but with the focus on the notion of categoricity as a prerequisite for physical theories. A first order theory is called categorical if it has a unique (up to isomorphism) model in every infinite cardinality. This is a very important insight for physics coming from model theory, and it can also be applied to characterize the choices of structures for mathematical models of nature, perhaps in the more permissive version of uncountable categoricity. Indeed, the complex Hilbert space is a totally categorical theory \cite{yaacov2008model,goldbring2012definable}.

\noindent 3. As we saw in the example of the qubit, there seems to be a connection between introducing hidden variables in a theory and having a non-trivial closure on the fixset lattice of its naming group. The connection is intriguing, because it might add a new perspective to the meaning of hidden variable completions, but its mechanism is not obvious. Hidden variable models for general quantum theories are quite subtle constructions that do not immediately fit in our algebraic framework \cite{bohm1966proposed,gudder1970hidden,jauch1968hidden,gudder1968hidden,bohm1968hidden}. From the point of view of group theory one can ask: Which permutation groups $G$ have the property $G<Aut(\mathcal{L}_{\mathrm{fix}}(G))$ and why?

\section*{Acknowledgements}
We would like to thank Eleni Vougiouklaki for providing insight into the linear representations of phase space automorphism groups, and Kostandis Loukos for clarifying certain philosophical questions.

\bibliographystyle{myunsrt}
\bibliography{refs}

\begin{thebibliography}{10}

\bibitem{logic_of_qm}
Garrett Birkhoff and John von Neumann.
\newblock {\em The logic of quantum mechanics}.
\newblock Annals of mathematics, pages 823--843, 1936.

\bibitem{von2018mathematical}
John~Von Neumann.
\newblock {\em {Mathematical foundations of quantum mechanics: New edition}},
  volume~53.
\newblock Princeton university press, 2018.

\bibitem{murray1936rings}
Francis~J Murray and John von Neumann.
\newblock {\em On rings of operators}.
\newblock Annals of Mathematics, pages 116--229, 1936.

\bibitem{murray1937rings}
Francis~J Murray and John von Neumann.
\newblock {\em {On rings of operators. II}}.
\newblock Transactions of the American Mathematical Society, 41(2):208--248,
  1937.

\bibitem{neumann1940rings}
John von Neumann.
\newblock {\em {On rings of operators. III}}.
\newblock Annals of Mathematics, pages 94--161, 1940.

\bibitem{murray1943rings}
Francis~J Murray and John von Neumann.
\newblock {\em {On rings of operators. IV}}.
\newblock Annals of Mathematics, pages 716--808, 1943.

\bibitem{araki1964type}
Huzihiro Araki.
\newblock {\em {Type of von Neumann algebra associated with free field}}.
\newblock Progress of Theoretical Physics, 32(6):956--965, 1964.

\bibitem{haag2012local}
Rudolf Haag.
\newblock {\em {Local quantum physics: Fields, particles, algebras}}.
\newblock Springer Science \& Business Media, 2012.

\bibitem{witten2022gravity}
Edward Witten.
\newblock {\em Gravity and the crossed product}.
\newblock Journal of High Energy Physics, 2022(10):1--28, 2022.

\bibitem{chandrasekaran2023algebra}
Venkatesa Chandrasekaran, Roberto Longo, Geoff Penington, and Edward Witten.
\newblock {\em {An algebra of observables for de Sitter space}}.
\newblock Journal of High Energy Physics, 2023(2):1--56, 2023.

\bibitem{stone1936theory}
Marshall~H Stone.
\newblock {\em {The theory of representation for Boolean algebras}}.
\newblock Transactions of the American Mathematical Society, 40(1):37--111,
  1936.

\bibitem{hodges1993model}
Wilfrid Hodges.
\newblock {\em Model theory}.
\newblock Cambridge University Press, 1993.

\bibitem{kochen1967problem}
Simon~B Kochen and Ernst Specker.
\newblock {\em The problem of hidden variables in quantum mechanics}.
\newblock J. Math. Mech., 17:59--87, 1967.

\bibitem{witten2022does}
Edward Witten.
\newblock {\em {Why Does Quantum Field Theory In Curved Spacetime Make Sense?
  And What Happens To The Algebra of Observables In The Thermodynamic Limit?}}
\newblock In {\em Dialogues Between Physics and Mathematics: CN Yang at 100},
  pages 241--284. Springer, 2022.

\bibitem{bergman2006closed}
George~M Bergman and Saharon Shelah.
\newblock {\em Closed subgroups of the infinite symmetric group}.
\newblock Algebra Universalis, 55(2-3):137--173, 2006.

\bibitem{mitchell2011generating}
James Mitchell, Michal Morayne, and Yann Peresse.
\newblock {\em Generating the infinite symmetric group using a closed subgroup
  and the least number of other elements}.
\newblock Proceedings of the American Mathematical Society, 139(2):401--405,
  2011.

\bibitem{jordan1871theoremes}
Camille Jordan.
\newblock {\em Th{\'e}oremes sur les groupes primitifs}.
\newblock Journal de Math{\'e}matiques Pures et Appliqu{\'e}es, 16:383--408,
  1871.

\bibitem{neumann1985some}
Peter~M Neumann.
\newblock {\em Some primitive permutation groups}.
\newblock Proceedings of the London Mathematical Society, 3(2):265--281, 1985.

\bibitem{neumann2006concept}
Peter~M Neumann.
\newblock {\em {The concept of primitivity in group theory and the second
  memoir of Galois}}.
\newblock Archive for history of exact sciences, 60(4):379--429, 2006.

\bibitem{kantor1985homogeneous}
William~M Kantor.
\newblock {\em Homogeneous designs and geometric lattices}.
\newblock Journal of combinatorial theory, series A, 38(1):66--74, 1985.

\bibitem{cherlin1985N0}
Gregory Cherlin, Leo Harrington, and Alistair~H Lachlan.
\newblock {\em $\aleph_0$-categorical, $\aleph_0$-stable structures}.
\newblock Annals of Pure and Applied Logic, 28(2):103--135, 1985.

\bibitem{mcdonough1981infinite}
Thomas~P McDonough.
\newblock {\em {An infinite version of Marggraff's theorem}}.
\newblock The Quarterly Journal of Mathematics, 32(2):173--179, 1981.

\bibitem{cameron1976transitivity}
Peter~J Cameron.
\newblock {\em Transitivity of permutation groups on unordered sets}.
\newblock Mathematische Zeitschrift, 148:127--139, 1976.

\bibitem{adeleke1996primitive}
Samson~Adepoju Adeleke and Peter~M Neumann.
\newblock {\em Primitive permutation groups with primitive Jordan sets}.
\newblock Journal of the London Mathematical Society, 53(2):209--229, 1996.

\bibitem{adeleke1996classification}
Samson~Adepoju Adeleke and Dugald Macpherson.
\newblock {\em Classification of infinite primitive Jordan permutation groups}.
\newblock Proceedings of the London Mathematical Society, 3(1):63--123, 1996.

\bibitem{bhattacharjee2006notes}
Meenaxi Bhattacharjee, R{\"o}gnvaldur~G M{\"o}ller, Dugald Macpherson, and
  Peter~M Neumann.
\newblock {\em Notes on infinite permutation groups}.
\newblock Springer, 2006.

\bibitem{beutelspacher1994transfinite}
Albrecht Beutelspacher and Peter~J Cameron.
\newblock {\em {Transfinite methods in geometry}}.
\newblock Bulletin of the Belgian Mathematical Society-Simon Stevin,
  1(3):337--347, 1994.

\bibitem{adeleke1996infinite}
Samson~Adepoju Adeleke.
\newblock {\em {Infinite Jordan Permutation Groups}}.
\newblock Ordered Groups and Infinite Permutation Groups, pages 159--194, 1996.

\bibitem{gleason1957measures}
Andrew~M Gleason.
\newblock {\em Measures on the Closed Subspaces of a Hilbert Space}.
\newblock Journal of Mathematics and Mechanics, pages 885--893, 1957.

\bibitem{hillman}
Chris Hillman.
\newblock {\em {An Outline of the Theory of G-Sets}}.
\newblock 1996.

\bibitem{hall2018theory}
Marshall Hall.
\newblock {\em The theory of groups}.
\newblock Courier Dover Publications, 2018.

\bibitem{GAP4}
The GAP~Group.
\newblock {\em {GAP -- Groups, Algorithms, and Programming, Version 4.12.2}},
  2022.

\bibitem{brazil1994maximal}
Marcus Brazil, Jacinta Covington, Tim Penttila, Cheryl~E Praeger, and Alan~R
  Woods.
\newblock {\em Maximal subgroups of infinite symmetric groups}.
\newblock Proceedings of the London Mathematical Society, 3(1):77--111, 1994.

\bibitem{macpherson1990subgroups}
Dugald Macpherson and Peter~M Neumann.
\newblock {\em Subgroups of infinite symmetric groups}.
\newblock Journal of the London Mathematical Society, 2(1):64--84, 1990.

\bibitem{macpherson1991large}
Dugald Macpherson.
\newblock {\em Large subgroups of infinite symmetric groups}.
\newblock Finite and infinite combinatorics in sets and logic (Banff, AB,
  1991), 411:249--278, 1991.

\bibitem{ball1966maximal}
Ralph~W Ball.
\newblock {\em Maximal subgroups of symmetric groups}.
\newblock Transactions of the American Mathematical Society, 121(2):393--407,
  1966.

\bibitem{rosenberg1958structure}
Alex Rosenberg.
\newblock {\em The structure of the infinite general linear group}.
\newblock Annals of Mathematics, pages 278--294, 1958.

\bibitem{macpherson1992maximal}
Dugald Macpherson.
\newblock {\em Maximal subgroups of infinite dimensional general linear
  groups}.
\newblock Journal of the Australian Mathematical Society, 53(3):338--351, 1992.

\bibitem{antoneli2006maximal}
Fernando Antoneli, Michael Forger, and Paola Gaviria.
\newblock {\em {Maximal subgroups of compact Lie groups}}.
\newblock arXiv preprint math/0605784, 2006.

\bibitem{marker2006model}
David Marker.
\newblock {\em Model theory: an introduction}, volume 217.
\newblock Springer Science \& Business Media, 2006.

\bibitem{sep-logic-firstorder-emergence}
William Ewald.
\newblock {\em {The Emergence of First-Order Logic}}.
\newblock In {\em The {Stanford} Encyclopedia of Philosophy}. Metaphysics
  Research Lab, Stanford University, 2019.

\bibitem{hanskamp}
Hans Kamp.
\newblock {\em {Lectures on Logic, Ch.III. Set Theory as a Theory of First
  Order Logic}}.

\bibitem{kadison1997fundamentals}
Richard~V Kadison and John~R Ringrose.
\newblock {\em {Fundamentals of the Theory of Operator Algebras}}, volume~2.
\newblock American Mathematical Society, 1997.

\bibitem{jonesrapid}
Vaughan Jones.
\newblock {\em {Lecture 1. von Neumann Algebras}}.

\bibitem{sep-qt-quantlog}
Alexander Wilce.
\newblock {\em {Quantum Logic and Probability Theory}}.
\newblock In {\em The {Stanford} Encyclopedia of Philosophy}. Metaphysics
  Research Lab, Stanford University, 2021.

\bibitem{turner2021ultrafilters}
Jason Turner.
\newblock {\em {Ultrafilters as Propositional Theories}}.
\newblock 2021.

\bibitem{birkhoff1940lattice}
Garrett Birkhoff.
\newblock {\em Lattice theory}, volume~25.
\newblock American Mathematical Society, 1940.

\bibitem{hawaii}
J.B. Nation.
\newblock {\em {Notes on Lattice Theory}}.

\bibitem{piron1976foundations}
Constantin Piron.
\newblock {\em On the foundations of quantum physics}.
\newblock Springer, 1976.

\bibitem{moore2009operational}
David~J Moore and Frank Valckenborgh.
\newblock {\em Operational quantum logic: a survey and analysis}.
\newblock Handbook of Quantum Logic and Quantum Structures--Quantum Logic,
  pages 389--441, 2009.

\bibitem{sep-logic-infinitary}
John~L Bell.
\newblock {\em {Infinitary Logic}}.
\newblock In {\em The {Stanford} Encyclopedia of Philosophy}. Metaphysics
  Research Lab, Stanford University, 2016.

\bibitem{yaacov2008model}
Ita{\"i}~Ben Yaacov, Alexander Berenstein, C.~Ward Henson, and Alexander
  Usvyatsov.
\newblock {\em Model theory for metric structures}.
\newblock London Mathematical Society Lecture Note Series, 350:315, 2008.

\bibitem{goldbring2012definable}
Isaac Goldbring.
\newblock {\em {Definable operators on Hilbert spaces}}.
\newblock Notre Dame Journal of Formal Logic, 53(2):193 -- 201, 2012.

\bibitem{yaacov2017metric}
Ita{\"\i}~Ben Yaacov, Michal Doucha, Andr{\'e} Nies, and Todor Tsankov.
\newblock {\em {Metric Scott analysis}}.
\newblock Advances in Mathematics, 318:46--87, 2017.

\bibitem{holland1995orthomodularity}
Samuel~S Holland.
\newblock {\em {Orthomodularity in infinite dimensions; a theorem of M.
  Sol\`{e}r}}.
\newblock Bulletin of the American Mathematical Society, 32(2):205--234, 1995.

\bibitem{bar2011wiener}
Christian B{\"a}r and Frank Pf{\"a}ffle.
\newblock {\em {Wiener measures on Riemannian manifolds and the Feynman-Kac
  formula}}.
\newblock arXiv preprint arXiv:1108.5082, 2011.

\bibitem{marsden1974reduction}
Jerrold Marsden and Alan Weinstein.
\newblock {\em {Reduction of symplectic manifolds with symmetry}}.
\newblock Reports on mathematical physics, 5(1):121--130, 1974.

\bibitem{halvorson2006algebraic}
Hans Halvorson and Michael M{\"u}ger.
\newblock {\em Algebraic quantum field theory}.
\newblock arXiv preprint math-ph/0602036, 2006.

\bibitem{witten2023algebras}
Edward Witten.
\newblock {\em {Algebras, Regions, and Observers}}.
\newblock arXiv preprint arXiv:2303.02837, 2023.

\bibitem{isono2019unique}
Yusuke Isono.
\newblock {\em {Unique prime factorization for infinite tensor product
  factors}}.
\newblock Journal of Functional Analysis, 276(7):2245--2278, 2019.

\bibitem{mcduff1970central}
Dusa McDuff.
\newblock {\em Central sequences and the hyperfinite factor}.
\newblock Proceedings of the London Mathematical Society, 3(3):443--461, 1970.

\bibitem{von1981continuous}
John von Neumann.
\newblock {\em Continuous geometries with a transition probability}, volume
  252.
\newblock American Mathematical Society, 1981.

\bibitem{redei1996john}
Mikl{\'o}s R{\'e}dei.
\newblock {\em {Why John von Neumann did not like the Hilbert space formalism
  of quantum mechanics (and what he liked instead)}}.
\newblock Studies In History and Philosophy of Science Part B: Studies In
  History and Philosophy of Modern Physics, 27(4):493--510, 1996.

\bibitem{ford1954homeomorphism}
Lester~R Ford.
\newblock {\em Homeomorphism groups and coset spaces}.
\newblock Transactions of the American Mathematical Society, 77(3):490--497,
  1954.

\bibitem{halvorson2001nature}
Hans Halvorson.
\newblock {\em On the nature of continuous physical quantities in classical and
  quantum mechanics}.
\newblock Journal of Philosophical Logic, 30:27--50, 2001.

\bibitem{adeleke1998relations}
Samson~Adepoju Adeleke and Peter~M Neumann.
\newblock {\em {Relations Related to Betweenness: Their Structure and
  Automorphisms}}, volume 623.
\newblock American Mathematical Society, 1998.

\bibitem{adeleke2013irregular}
Samson~Adepoju Adeleke.
\newblock {\em {On irregular infinite Jordan groups}}.
\newblock Communications in Algebra, 41(4):1514--1546, 2013.

\bibitem{bhattacharjee2006jordan}
Meenaxi Bhattacharjee and Dugald Macpherson.
\newblock {\em Jordan groups and limits of betweenness relations}.
\newblock 2006.

\bibitem{almazaydeh2022jordan}
Asma~Ibrahim Almazaydeh and Dugald Macpherson.
\newblock {\em {Jordan permutation groups and limits of D-relations}}.
\newblock Journal of Group Theory, 25(3):447--508, 2022.

\bibitem{bradley2023limits}
David Bradley-Williams and John~K Truss.
\newblock {\em On limits of betweenness relations}.
\newblock Journal of Group Theory, 2023.

\bibitem{almazaydeh2019infinite}
Asma~Ibrahim Almazaydeh.
\newblock {\em {Infinite Jordan permutation groups}}.
\newblock PhD thesis, University of Leeds, 2019.

\bibitem{adeleke1995semilinear}
Samson~Adepoju Adeleke.
\newblock {\em {Semilinear tower of Steiner systems}}.
\newblock Journal of Combinatorial Theory, Series A, 72(2):243--255, 1995.

\bibitem{johnson2002constructions}
Keith Johnson.
\newblock {\em {Constructions of semilinear towers of Steiner systems}}.
\newblock Tits buildings and the model theory of groups (Ed. K. tent), London
  Math. Soc. Lecture Notes, 291:235--278, 2002.

\bibitem{hrushovski1993new}
Ehud Hrushovski.
\newblock {\em A new strongly minimal set}.
\newblock Annals of pure and applied logic, 62(2):147--166, 1993.

\bibitem{cameron2002infinite}
Peter~J Cameron and Bridget~S Webb.
\newblock {\em What is an infinite design?}
\newblock Journal of Combinatorial Designs, 10(2):79--91, 2002.

\bibitem{pitowsky1998infinite}
Itamar Pitowsky.
\newblock {\em {Infinite and finite Gleason’s theorems and the logic of
  indeterminacy}}.
\newblock Journal of Mathematical Physics, 39(1):218--228, 1998.

\bibitem{aaronson2004quantum}
Scott Aaronson.
\newblock {\em Is quantum mechanics an island in theoryspace?}
\newblock arXiv preprint quant-ph/0401062, 2004.

\bibitem{kapustin2013there}
Anton Kapustin.
\newblock {\em {Is there life beyond Quantum Mechanics?}}
\newblock arXiv preprint arXiv:1303.6917, 2013.

\bibitem{wootters1981statistical}
William~K Wootters.
\newblock {\em {Statistical distance and Hilbert space}}.
\newblock Physical Review D, 23(2):357, 1981.

\bibitem{mackey1963mathematical}
George~W Mackey.
\newblock {\em {Mathematical Foundations of Quantum Mechanics}}.
\newblock W. A. Benjamin, Inc., New York, 1963.

\bibitem{piron1964axiomatique}
Constantin Piron.
\newblock {\em Axiomatique quantique}.
\newblock Helvetica physica acta, 37(4-5):439, 1964.

\bibitem{soler1995characterization}
Maria~Pia Sol\`{e}r.
\newblock {\em {Characterization of Hilbert spaces by orthomodular spaces}}.
\newblock Communications in Algebra, 23(1):219--243, 1995.

\bibitem{zilber2014perfect}
Boris Zilber.
\newblock {\em Perfect infinities and finite approximation}.
\newblock In {\em Infinity and truth}, pages 199--223. World Scientific, 2014.

\bibitem{cruz2015geometric}
John A~Cruz Morales and Boris Zilber.
\newblock {\em The geometric semantics of algebraic quantum mechanics}.
\newblock Philosophical Transactions of the Royal Society A,
  373(2047):20140245, 2015.

\bibitem{cruz2021around}
John A~Cruz Morales, Andr{\'e}s Villaveces, and Boris Zilber.
\newblock {\em Around logical perfection}.
\newblock Theoria, 87(4):971--985, 2021.

\bibitem{bohm1966proposed}
David Bohm and Jeffrey Bub.
\newblock {\em A proposed solution of the measurement problem in quantum
  mechanics by a hidden variable theory}.
\newblock Reviews of Modern Physics, 38(3):453, 1966.

\bibitem{gudder1970hidden}
Stanley~P Gudder.
\newblock {\em {On Hidden-Variable Theories}}.
\newblock Journal of Mathematical Physics, 11(2):431--436, 1970.

\bibitem{jauch1968hidden}
Joseph-Maria Jauch and Constantin Piron.
\newblock {\em Hidden variables revisited}.
\newblock Reviews of Modern Physics, 40(1):228, 1968.

\bibitem{gudder1968hidden}
Stanley~P Gudder.
\newblock {\em Hidden variables in quantum mechanics reconsidered}.
\newblock Reviews of Modern Physics, 40(1):229, 1968.

\bibitem{bohm1968hidden}
David Bohm and Jeffrey Bub.
\newblock {\em {On hidden variables—A reply to comments by Jauch and Piron
  and by Gudder}}.
\newblock Reviews of Modern Physics, 40(1):235, 1968.

\end{thebibliography}

\end{document}